\documentclass{edp-jp4}
\usepackage{graphicx}
\begin{document}

%%-----------------------------
%%      the top matter
%%-----------------------------
\title{Potentials and distribution functions to be used for dynamical modeling with GAIA-like data}\thanks{we would like to thank A.Jorissen an K.Van Caelenberg for their precious help.} 
\author{B. Famaey}\address{Institut d'Astronomie et d'Astrophysique, Universit\'e Libre de Bruxelles, Belgium; Ph.D. student F.R.I.A.; \email{bfamaey@astro.ulb.ac.be}}
\author{H. Dejonghe}\address{Sterrenkundig Observatorium, Universiteit Gent, Belgium}
\maketitle
%
%%-----------------------------
%%      your text
%%-----------------------------
\section{Introduction}
The GAIA mission will offer the wonderful opportunity to establish galactic dynamical models based on the data of a large number of stars. Indeed, complete positional and kinematical data over a large volume in the Galaxy are needed to construct well constrained models: we do not have such data for the moment, but this is precisely what the GAIA satellite will provide.

We present here new tools to establish axisymmetric equilibrium models of the Milky Way, based on GAIA-like data. We know that our Galaxy is a barred one and so, the axisymmetric hypothesis is not correct, but axisymmetric models are a prerequisite for perturbation theory (for example the theory of spiral density waves) and thus they are still very useful.

The models we wish to establish are pairs $(V,F)$ where $V$ is the gravitational potential generated by the whole mass distribution including the dark matter, and $F$ is the distribution function in phase space for late-type tracer stars in the galactic disk. For an equilibrium model (stationary potential and stationary distribution function solution of the collisionless Boltzmann equation), we know that the distribution function in phase space depends only on the integrals of the motion. During the last decade, many studies (e.g. \cite{Du96}) have shown that the distribution function of tracer stars in the Milky Way has to depend on three isolating integrals of the motion, especially if information on the vertical motion of the stars is present as it will be with GAIA. In order to have an analytic third integral, in addition to the binding energy and the vertical component of the angular momentum, we use St\"ackel potentials \cite{Z85}.

\section{The potentials}
In this contribution, we continue the work of Batsleer \& Dejonghe \cite{Ba94}, who presented a set of simple axisymmetric St\"ackel potentials with two mass components (halo and disk) and a flat rotation curve. The existence of a thick disk as a separate stellar component is now well documented (e.g. \cite{Oh94}): so we have generalized the Batsleer's potentials by adding a thick disk to them. Our new St\"ackel potentials are described by five parameters. They turn out to have an effective bulge because the mass density grows faster than an exponential near the center in the plane. The other motivation to continue Batsleer's work is that, in recent years, Hipparcos data have enabled an accurate determination of some fundamental galactic parameters in the solar neighbourhood: the mass density$ \rho_{\odot}$ \cite{Cr98,Ho00} and the Oort constants $A$ and $B$ \cite{Fe97}. We have looked for the St\"ackel parameters that are consistent with the above observables: as can be seen on Fig.1, many different combinations of the parameters are found to be consistent with those observables.
\begin{figure}
\begin{center}
 \includegraphics[width=4.1cm,angle=-90]{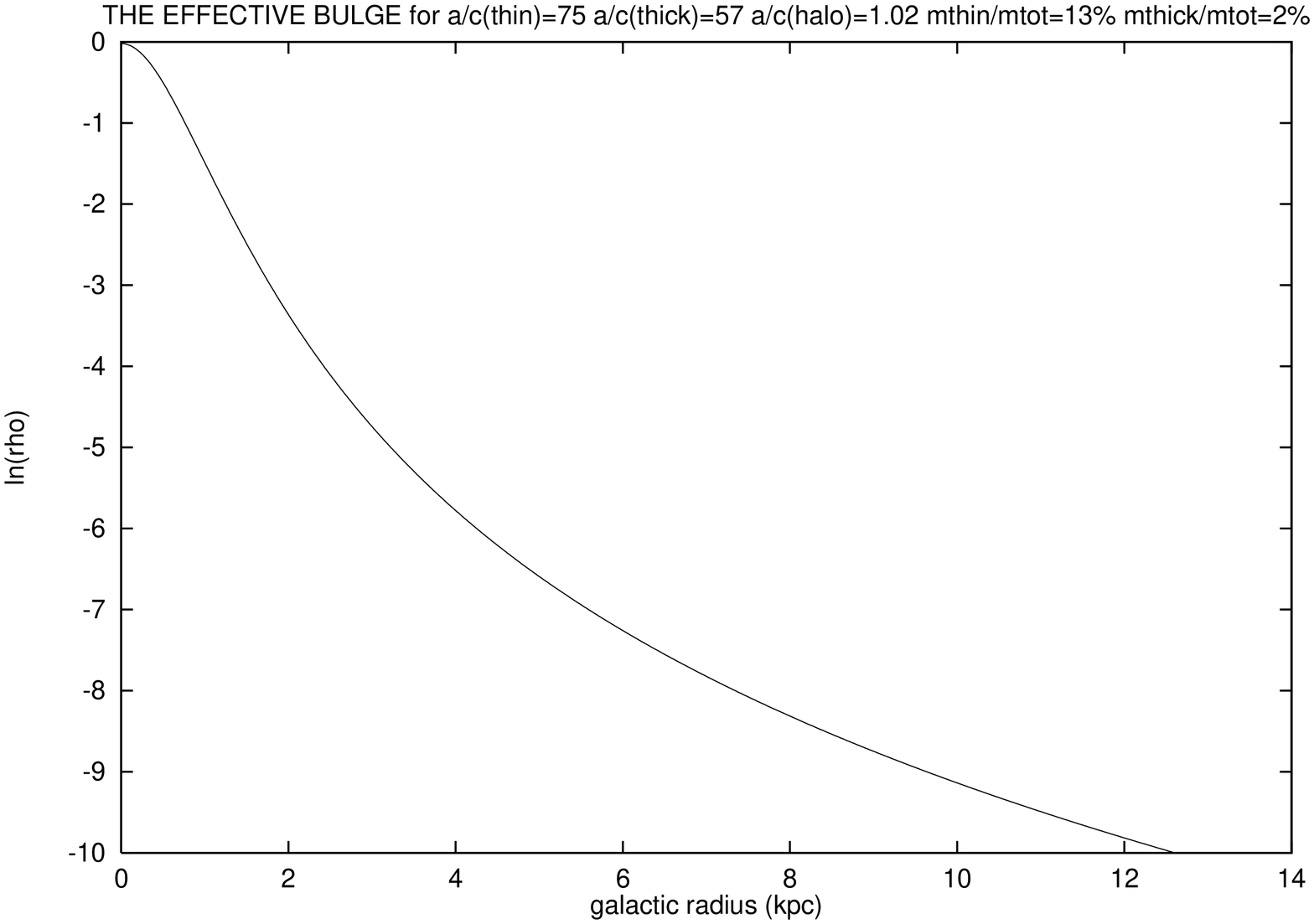}
 \qquad
 \includegraphics[width=4.1cm,angle=-90]{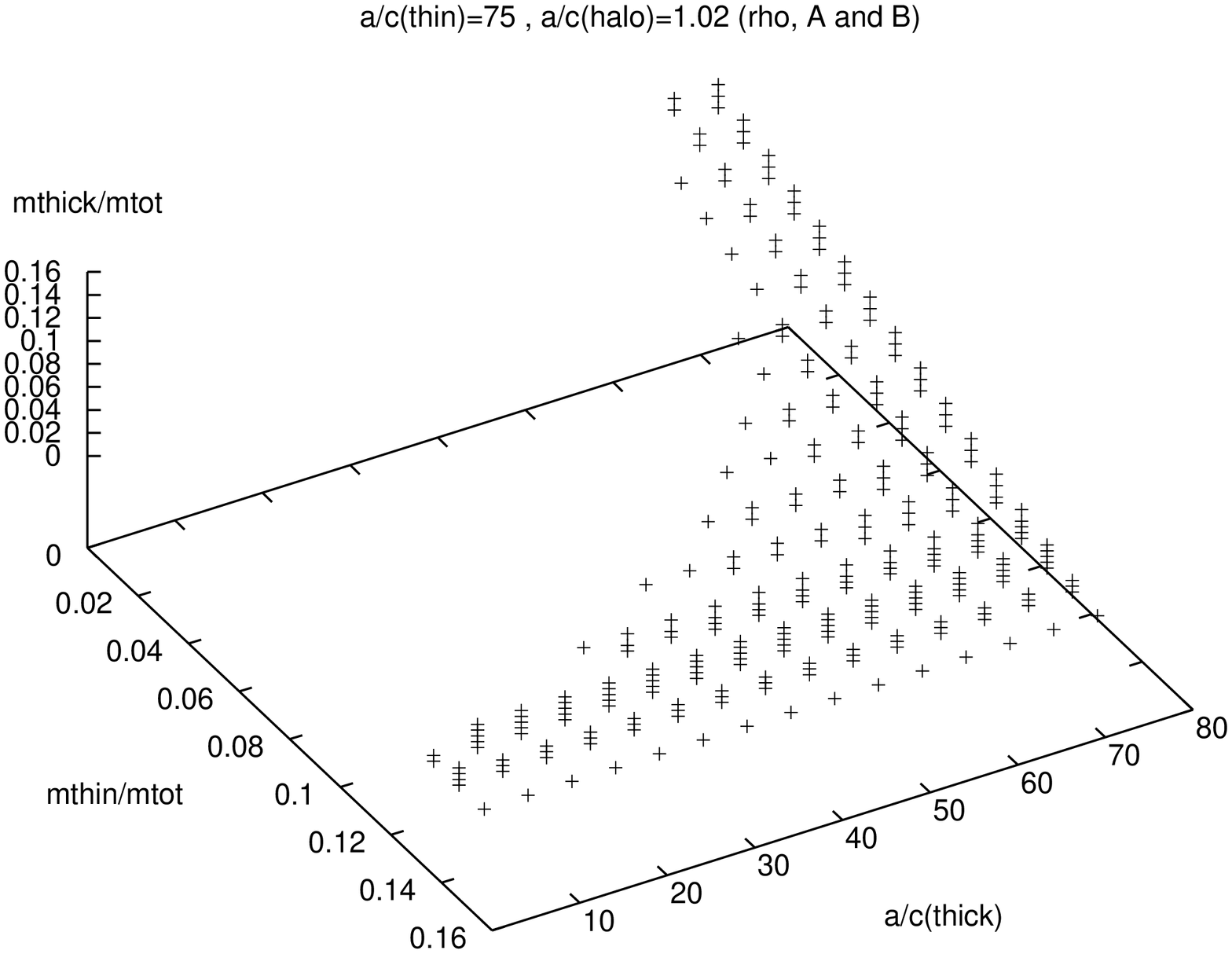}
 \caption{Left panel: {\footnotesize The mass density in the plane.} Right panel: {\footnotesize For a fixed value of the axis ratio of the thin disk and of the halo, the figure displays all the values of the axis ratio of the thick disk and of the contributions of the disks to the total mass that are consistent with Hipparcos observables.}}
\end{center}
\end{figure}

\section{The modified Fricke components and the modeling}
In the future we will test the ability of these potentials to be combined with a distribution function, in order to reproduce kinematical data like those that will be provided by GAIA. This distribution function will be expressed as a linear combination of basis functions depending on a few parameters.
We have defined new component distribution functions (modified Fricke components) with 9 parameters, {\it that depend on three integrals of the motion}, and that can represent realistic stellar disks when a judicious linear combination of them is chosen in a realistic galactic potential. These components have a finite extent in the $z$-direction and have realistic scale lengths. More details can be found in \cite{Fa01}.

To construct our model, the method will be iterative: we will choose a potential, find a linear combination of modified Fricke components that fits the data by using the quadratic programming technique described in \cite{De89}, and then modify the potential in the light of the quality of that fit. This will provide some new constraints on the mass distribution in the Galaxy and some information on the dynamical state of the different late-type stars (since the Galaxy is a system with a long memory, this could provide information on the history of the Milky Way).

%%-----------------------------
%%      your bibliography
%%-----------------------------

\end{document}